\documentclass[aps,prb,reprint,superscriptaddress]{revtex4-2}
\usepackage{graphicx}
\usepackage{amsmath}
\usepackage{amssymb}
\usepackage{bm}
\usepackage{physics}
\usepackage{subfigure}
\usepackage[colorlinks=true,linkcolor=blue,urlcolor=blue,citecolor=blue]{hyperref}
\usepackage{times}
\usepackage{changes}
\begin{document}

\newcommand{\Hop}{\hat{H}}
\newcommand{\Himp}{\hat{H}_{\rm imp}}
\newcommand{\Uimp}{\hat{U}_{\rm imp}}
\newcommand{\gHimp}{\mathcal{H}_{\rm imp}}
\newcommand{\gUimp}{\mathcal{U}_{\rm imp}}
\newcommand{\Hbath}{\hat{H}_{\rm bath}}
\newcommand{\Hhyb}{\hat{H}_{\rm hyb}}

\newcommand{\aop}{\hat{a}}
\newcommand{\adop}{\hat{a}^{\dagger}}
\newcommand{\sgp}{\hat{\sigma}^+}
\newcommand{\sgx}{\hat{\sigma}^x}
\newcommand{\sgy}{\hat{\sigma}^y}
\newcommand{\sgz}{\hat{\sigma}^z}
\newcommand{\nop}{\hat{n}}

\newcommand{\cop}{\hat{c}}
\newcommand{\cdop}{\hat{c}^{\dagger}}
\newcommand{\hc}{{\rm H.c.}}
\newcommand{\rhotot}{\hat{\rho}_{\mathrm{tot}}}
\newcommand{\rhoop}{\hat{\rho}}
\newcommand{\rhoimp}{\hat{\rho}_{\mathrm{imp}}}
\newcommand{\rhobath}{\hat{\rho}_{\mathrm{bath}}}
\newcommand{\Zimp}{Z_{{\rm imp}}}
\newcommand{\mea}{\mathcal{D}}
\newcommand{\gK}{\mathcal{K}}
\newcommand{\gI}{\mathcal{I}}
\newcommand{\bolda}{\bm{a}}
\newcommand{\boldabar}{\bar{\bm{a}}}
\newcommand{\abar}{\bar{a}}
\newcommand{\im}{{\rm i}}
\newcommand{\contour}{\mathcal{C}}
\newcommand{\gA}{\mathcal{A}}
\newcommand{\gB}{\mathcal{B}}
\newcommand{\gM}{\mathcal{M}}
\newcommand{\boldeta}{\bm{\eta}}
\newcommand{\boldetabar}{\bar{\bm{\eta}}}
\newcommand{\etabar}{\bar{\eta}}
\newcommand{\parity}{\mathcal{P}}
\newcommand{\current}{\mathcal{J}}

\newcommand{\mse}{\mathcal{E}}

\newcommand{\xs}[1]{{(\color{red}#1)}}
\newcommand{\gcc}[1]{{\color{black}#1}}
\definechangesauthor[name=RF,color=blue]{RF}

% \title{Solving Equilibrium Impurity Problems in Real-Frequency axis Using Grassmann Time-Evolving Matrix Product Operators}
\title{Real-time Impurity Solver Using Grassmann Time-Evolving Matrix Product Operators}
\author{Ruofan Chen}
\affiliation{College of Physics and Electronic Engineering, and Center for Computational Sciences, Sichuan Normal University, Chengdu 610068, China}
\author{Xiansong Xu}
\affiliation{College of Physics and Electronic Engineering, and Center for Computational Sciences, Sichuan Normal University, Chengdu 610068, China}
\affiliation{Science and Math Cluster, Singapore University of Technology and Design, 8 Somapah Road, Singapore 487372}
\author{Chu Guo}
\email{guochu604b@gmail.com}
\affiliation{Key Laboratory of Low-Dimensional Quantum Structures and Quantum Control of Ministry of Education, Department of Physics and Synergetic Innovation Center for Quantum Effects and Applications, Hunan Normal University, Changsha 410081, China}
\date{\today}

\begin{abstract}
An emergent and promising tensor-network-based impurity solver is to represent the Feynman-Vernon influence functional as a matrix product state, where the bath is integrated out analytically.
Here we present an approach to calculate the equilibrium impurity spectral function based on the recently proposed Grassmann time-evolving matrix product operators method. The central idea is to perform a quench from a separable impurity-bath initial state as in the non-equilibrium scenario. The retarded Green's function $G(t+t_0, t'+t_0)$ is then calculated after an equilibration time $t_0$ such that the impurity and bath are approximately in thermal equilibrium. There are two major advantages of this method. First, since we focus on real-time dynamics, we do not need to perform the numerically ill-posed analytic continuation as in the imaginary time evolution based methods. Second, the required bond dimension of the matrix product state in real-time calculations is observed to be much smaller than that in imaginary-time calculations, leading to a significant improvement in numerical efficiency.
The accuracy of this method is demonstrated in the single-orbital Anderson impurity model and benchmarked against the continuous-time quantum Monte Carlo method.
\end{abstract}
\maketitle

\section{Introduction}
% Anderson impurity model \cite{anderson1961-localized}.

% Kondo model and fermionic open systems \cite{hewson1993-the,Wilson1975}

% Review of DMFT \cite{GeorgesRozenberg1996}.

% Path integral \cite{FeynmanVernon1963}.

% Grassmann path integral
% \cite{negele1998-quantum,kamenev2009-keldysh}.

% Review of continuous time monte-carlo \cite{GullWerner2011}.

% QUAPI \cite{makarov1994-path,makri1995-numerical,dattani2012-analytic}.

% MPS and MPO
% \cite{schollwoeck2005-density,schollwoeck2011-density,orus2014-practical}

% TEMPO \cite{StrathearnLovett2018,strathearn2020-modelling,gribben2022-exact}

% Imaginary time TEMPO \cite{chiu2022-numerical}

% Other Grassmann tensor network \cite{gu2013-efficient,yoshimura2018-calculation,akiyama2021-more}.

% Fishman-White \cite{fishman2015-compression}

% MPS real time DMFT \cite{GanahlVerstraete2015,kohn2021-efficient}.

% MPS imaginary time DMFT \cite{WolfSchollwock2015,BauernfeindEvertz2017}.

% Current \cite{chen2023-heat}.

% Abanin \cite{ThoennissAbanin2023a,ThoennissAbanin2023b}.

% Abanin imaginary time \cite{KlossAbanin2023}.

% A. J. Millis and D. R. Reichman \cite{ng2023-real}.
The dynamical mean-field theory (DMFT) is one of the most successful numerical methods for strongly correlated electron systems beyond one dimension~\cite{MetznerVollhardt1989,GeorgesKotliar1992,GeorgesRozenberg1996,KotliarMarianetti2006}. The essential idea of DMFT is a mapping of lattice models into self-consistent quantum impurity models. The crucial step during the self-consistency
DMFT loop is to solve the 
quantum impurity problem (QIP) and obtain the impurity spectral function.
% or more concretely, to calculate the impurity spectral function.
Various approaches have been developed to solve QIPs, including the continuous-time quantum Monte Carlo (CTQMC)~\cite{GullWerner2011,RubtsovLichtenstein2005,GullTroyer2008,WernerMillis2006b,WernerMillis2006,ShinaokaWerner2017,EidelsteinCohen2020}, the numeric renormalization group (NRG)~\cite{Wilson1975,Bulla1999,BullaPruschke2008,Frithjof2008,ZitkoPruschke2009,DengGeorges2013,LeeWeichselbaum2016,LeeWeichselbaum2017}, exact diagonalization~\cite{CaffarelKrauth1994,KochGunnarsson2008,GranathStrand2012,LuHaverkort2014,ZaeraLin2020}, and matrix product state (MPS) based methods~\cite{WolfSchollwock2014b,GanahlEvertz2014,GanahlVerstraete2015,WolfSchollwock2015,GarciaRozenberg2004,NishimotoJeckelmann2006,WeichselbaumDelft2009,BauernfeindEvertz2017,WernerArrigoni2023,KohnSantoro2021,KohnSantoro2022}.
Among these approaches, the CTQMC methods have been extremely powerful in calculating the Green's function in the imaginary-time axis (the Matsubara Green's function)~\cite{WernerMillis2006b,Haule2007,WernerMillis2007,WernerMillis2008,ChanMillis2009}, where the spectral function can be obtained via analytic continuation.
However, these methods could suffer from the sign problem~\cite{TroyerWiese2005} and the analytic continuation is also numerically ill-posed for Monte Carlo results~\cite{WolfSchollwock2015,FeiGull2021}. 
Besides the CTQMC methods, the remaining methods directly store the impurity-bath wave function by explicitly discretizing the bath into a finite number of fermionic modes. Although they can work in both real-time and imaginary-time axes, their numerical efficiency is significantly hampered by the explicit treatment of the bath.
The NRG methods, in particular, have been used for real-time calculations of QIPs with up to three orbitals~\cite{MitchellBulla2014,StadlerWeichselbaum2015,HorvatMravlje2016,KuglerGeorges2020}. However, in general, these methods may either lack scalability to higher-orbital QIPs or lack control over the errors induced by bath discretization. 

A promising tensor-network-based impurity solver emerging in recent years is to make use of the Feynman-Vernon influence functional (IF) to analytically integrate out the bath and represent the multi-time impurity dynamics as an MPS in the temporal domain~\cite{StrathearnLovett2018,ThoennissAbanin2023a,ThoennissAbanin2023b,NgReichman2023,KlossAbanin2023,ChenGuo2023,ChenGuo2024b}. 
Such approaches are thus free of bath discretization errors and potentially more efficient than the conventional MPS-based methods.
They have been applied to study bosonic impurity problems~\cite{joergensen2019-exploiting,popovic2021-quantum,fux2021-efficient,gribben2021-using,otterpohl2022-hidden,gribben2022-exact} and fermionic impurity problems in both the non-equilibrium~\cite{ThoennissAbanin2023a,ThoennissAbanin2023b,NgReichman2023,ChenGuo2023} and imaginary-time equilibrium scenarios~\cite{KlossAbanin2023,ChenGuo2024b}. The Grassmann time-evolving matrix product operator (GTEMPO) method proposed in Ref.~\cite{ChenGuo2023} presents an efficient construction of the fermionic path integral (PI) as Grassmann MPSs (GMPSs) which respect the anti-commutation relation.
% , analogous to its bosonic counterpart~\cite{StrathearnLovett2018}. 
Since GTEMPO only relies on the PI, it can be straightforwardly used for real-time, imaginary-time or even mixed-time calculations. 
Therefore, in principle, it can be used to directly calculate the equilibrium retarded Green's function with a mixed-time PI expression. However, for GTEMPO it has been observed that larger bond dimensions are required in imaginary-time calculations than in real-time calculations~\cite{ChenGuo2024b} (the situation seems to be similar for the tensor network IF method in Ref.~\cite{KlossAbanin2023}), which is in sharp comparison with observations in the conventional MPS-based approaches~\cite{WolfSchollwock2015}.

In this work, we propose the use of GTEMPO method to calculate the spectral function, which is purely based on the non-equilibrium real-time dynamics of the impurity model from a separable impurity-bath initial state.
The central idea is to wait for a fixed time $t_0$ such that the impurity and bath equilibrate, and then calculate the retarded Green's function $G(t+t_0, t'+t_0)$.
% Similar ideas have been explored with wave function based methods (See Refs.\cite{BauerTroyer2016,KohnSantoro2021,KohnSantoro2022} for example), while our approach is built on top of the GTEMPO method. 
In this way, we can reuse all the algorithms in the non-equilibrium GTEMPO method and avoid imaginary-time calculations with GTEMPO completely. 
The only three hyperparameters in this approach, are the time discretization $\delta t$, the MPS bond dimension $\chi$ and the equilibration time $t_0$.
% Compared to imaginary-time calculations, this method primarily offers two advantages: 1) it eliminates the need for analytic continuation, and 2) it exhibits a significantly slower growth in the bond dimension of the Matrix Product States (MPS) during real-time calculations.
Compared to imaginary-time calculations~\cite{KlossAbanin2023,ChenGuo2024b}, the advantage of this method is mainly two-fold: 1) it does not require analytic continuation, and 2) it exhibits a much slower growth of the bond dimension of MPS for real-time calculations as compared to that in imaginary-time calculations.
% We demonstrate the advantage of this approach over the imaginary-time GTEMPO method by showing that a much smaller bond dimension of the MPS is required in this approach compared to the latter. 
The accuracy of this approach is benchmarked against CTQMC calculations for the single-orbital Anderson impurity model (AIM) for a wide range of interaction strengths.
Our work thus opens the door to practical applications of the GTEMPO method as a real-time quantum impurity solver.

\section{The non-equilibrium GTEMPO method}

% We first briefly review the non-equilibrium GTEMPO method \cite{ChenGuo2023} for the real-time dynamics of QIPs starting from a separable impurity-bath initial state, which is the building block for us to calculate the spectral function.
In this section, we present the formulation of the non-equilibrium GTEMPO method~\cite{ChenGuo2023}. We first give a basic recap on the quantum impurity model and several useful notations. 
Then we introduce the real-time path integral formalism in terms of Grassmann variables (GVs), as well as the definitions of the non-equilibrium Green's functions based on the Grassmann path integral.
% The real-time dynamics of path integral based on Grassmann variables and the corresponding non-equilibrium Green's functions calculations are then introduced. 

%  starting from a separable impurity-bath initial state, which is the building block for us to calculate the spectral function.

\subsection{Quantum impurity Hamiltonians}
The Hamiltonian for a general QIP can be written as 
\begin{align}
\Hop = \Himp + \Hbath + \Hhyb,
\end{align}
where $\Himp$ is the impurity Hamiltonian usually given by 
\begin{align}\label{eq:Himp}
\Himp = \sum_{p, q} t_{p, q} \adop_{p}\aop_{q} + \sum_{p,q,r,s} v_{p,q,r,s} \adop_p\adop_q\aop_r\aop_s .
\end{align}
Here $\adop$, $\aop$ are the fermionic creation and annihilation operators. The subscripts $p,q,r,s$ represent the fermion flavors that contain both the spin and orbital indices.
$\Hbath$ is the bath Hamiltonian which can be written as
\begin{align}
\Hbath = \sum_{p, k} \varepsilon_{p, k} \cdop_{p, k} \cop_{p, k}, 
\end{align}
where $k$ denotes the momentum label, and $\varepsilon_{p, k}$ denotes corresponding energy.  $\Hhyb$ is the hybridization Hamiltonian between the impurity and bath. We will focus on linear coupling such that the bath can be integrated out using the Feynman-Vernon IF~\cite{FeynmanVernon1963,negele1998-quantum,kamenev2009-keldysh}, which takes the following form:
\begin{align}
\Hhyb = \sum_{p, k} V_{p, k}\left(\adop_{p} \cop_{p, k} + \cdop_{p, k}\aop_{p} \right), 
\end{align}
with $V_{p, k}$ indicating the hybridization strength. For notational simplicity and as the general practice, we focus on flavor-independent $\Hbath$ and $\Hhyb$, such that we can write $\varepsilon_{p, k}=\varepsilon_{k}$ and $V_{p, k}=V_{k}$ for brevity.

\subsection{Notations and definitions}

Now we introduce several notations that will be used throughout this work. The separable impurity-bath initial state is denoted as 
\begin{align}\label{eq:separable_state}
\rhotot^{{\rm neq}} = \rhoimp \otimes \rhobath^{{\rm eq}},
\end{align}
where $\rhoimp$ is some arbitrary impurity state and $\rhobath^{{\rm eq}} \propto e^{-\beta \Hbath}$ is the thermal equilibrium state of the bath at inverse temperature $\beta$. The thermal equilibrium state of the impurity plus bath at inverse temperature $\beta$ will be denoted as
\begin{align}\label{eq:thermal_state}
\rhotot^{{\rm eq}} \propto e^{-\beta\Hop}.
\end{align}
The non-equilibrium retarded Green's function will be denoted as (here $\Theta(t)$ is the Heaviside step function)
\begin{align}
G^{{\rm neq}}_{p,q}(t, t') = \Theta(t-t')[G^{{\rm neq}, >}_{p,q}(t, t') - G^{{\rm neq}, <}_{p,q}(t, t') ].
\end{align}
Here $G^{{\rm neq}, >}_{p,q}(t, t')$ and $G^{{\rm neq}, <}_{p,q}(t, t')$ denote the greater and lesser Green's functions, respectively. They are defined as  
\begin{align}
\im G^{{\rm neq}, >}_{p,q}(t, t') &= \expval*{\aop_p(t)\adop_q(t')}_{\mathrm{neq}}\label{eq:neq_greater};\\
\im G^{{\rm neq}, <}_{p,q}(t, t') &= -\expval*{\adop_q(t') \aop_p(t)}_{\mathrm{neq}}\label{eq:neq_lesser},
\end{align}
where $\expval*{\cdots}_{\mathrm{neq}}=\Tr[\rhotot^{\mathrm{neq}}\cdots]/\Tr[\rhotot^{\mathrm{neq}}]$, $\aop(t) = e^{\im \Hop t} \aop e^{-\im \Hop t}$ and $\adop(t) = e^{\im \Hop t} \adop e^{-\im \Hop t}$.
The subscript ${\rm neq}$ in $\expval*{\cdots}_{\mathrm{neq}}$ stresses that these Green's functions are calculated with respect to the separable initial state $\rhotot^{{\rm neq}}$. Similarly, the equilibrium retarded Green's function will be denoted as 
\begin{align}
G_{p,q}(t, t') = \Theta(t-t')[G^{>}_{p,q}(t, t') - G^{<}_{p,q}(t, t') ],
\end{align}
with
\begin{align}
\im G^{>}_{p,q}(t, t') &= \expval*{\aop_p(t)\adop_q(t')}_{\mathrm{eq}}\label{eq:greater} ; \\
\im G^{<}_{p,q}(t, t') &= -\expval*{\adop_q(t') \aop_p(t)}_{\mathrm{eq}}\label{eq:lesser},
\end{align}
where $\expval*{\cdots}_{\mathrm{eq}}=\Tr[\rhotot^{\mathrm{eq}}\cdots]/\Tr[\rhotot^{\mathrm{eq}}]$.
$G_{p,q}(t, t')$ is time-translationally invariant, i.e., $G_{p,q}(t, t') = G_{p,q}(t-t')$. Therefore we can obtain the retarded Green's function in the real-frequency domain as
\begin{align} \label{eq:fourier}
G_{p,q}(\omega)=\int_0^{\infty}e^{\im(\omega+\im0) t}G_{p,q}(t)\dd{t},
\end{align}
where $\im0$ is an infinitesimal imaginary number.
The real-frequency Green's function with same flavor, $G_{p}(\omega) =G_{p,p}(\omega)$, is of particular interest, with which the spectral function $A_p(\omega)$ can be calculated as
\begin{align}\label{eq:spectrum_function}
A_p(\omega) = -\frac{1}{\pi} {\rm Im}\left[G_{p}(\omega)\right] .
\end{align}

This spectral function establishes a connection between the equilibrium retarded Green's function and the Matsubara Green's function. The Matsubara Green's function is defined in the imaginary-time axis as
% \begin{equation}
%   G_{pq}(\tau,\tau')=
%   \begin{cases}
%     -\Tr[\rhotot^{\mathrm{eq}}\aop_p(\tau)\adop_q(\tau')]/\Tr[\rhotot^{\mathrm{eq}}], & \tau\ge\tau';\\
%     \Tr[\rhotot^{\mathrm{eq}}\adop_q(\tau')\aop_p(\tau)]/\Tr[\rhotot^{\mathrm{eq}}], & \tau<\tau',\\
%   \end{cases}
% \end{equation}
\begin{align}
  \mathcal{G}_{pq}(\tau,\tau')=
  \begin{cases}
    -\expval*{\aop_p(\tau)\adop_q(\tau')}_{\mathrm{eq}}, & \tau\ge\tau';\\
    \expval*{\adop_q(\tau') \aop_p(\tau)}_{\mathrm{eq}}, & \tau<\tau',\\
  \end{cases}
\end{align}
where $\aop_p(\tau)=e^{\tau\Hop}\aop_pe^{-\tau\Hop}$ and $\adop_p(\tau)=e^{\tau\Hop}\adop_pe^{-\tau\Hop}$. The Matsubara Green's function is time-translationally invariant for $-\beta\le\tau-\tau'\le\beta$, and is anti-periodic that $\mathcal{G}_{pq}(\tau)=-\mathcal{G}_{pq}(\tau+\beta)$ for $\tau<0$. Thus it can be expanded as a Fourier series over range $0\le\tau\le\beta$ that
\begin{align}
  \mathcal{G}_{pq}(\tau)=\frac{1}{\beta}\sum_{m=-\infty}^{\infty}\mathcal{G}_{pq}(\im\omega_m)e^{-\im\omega_m\tau},
\end{align}
where $\omega_m=(2m+1)\pi/\beta$ and $\mathcal{G}_{pq}(\im\omega_m)$ is the imaginary-frequency Green's function
\begin{align}
  \mathcal{G}_{pq}(\im\omega_m)=\int_0^{\beta}e^{\im\omega_m\tau}\mathcal{G}_{pq}(\tau)\dd{\tau}.
\end{align}

The real- and imaginary-frequency Green's functions with the same flavor can be expressed by spectral functions as \cite{lifshitz1980-statistical,mahan2000-many}
\begin{align}
  G_p(\omega)&=\int\frac{A_p(\varepsilon)}{\omega+i0-\varepsilon}\dd{\varepsilon}, \\
  \mathcal{G}_p(\im\omega_m)&=\int\frac{A_p(\varepsilon)}{\im\omega_{m}-\varepsilon}\dd{\varepsilon},\label{eq:Giw}
\end{align}
which means that they are the same function in the complex plane:
\begin{equation}
G_p(z)=\int\frac{A_p(\varepsilon)}{z-\varepsilon}\dd{\varepsilon},
\end{equation}
but along the real and imaginary axis respectively. 

The Green's functions $G_p(t),G_p(\omega),\mathcal{G}_p(\tau),\mathcal{G}_p(\im \omega_m)$ can be obtained once the spectral function $A_p(\omega)$ is known. Thus determining the spectral function is crucial as it conveys the essential information of these Green's functions. The spectral function $A_p(\omega)$ can be easily extracted from the real-frequency Green's function via Eq.\eqref{eq:spectrum_function}. However, to obtain it from imaginary-time or frequency Green's function via inverting Eq.\eqref{eq:Giw} is numerically ill-posed: small fluctuations in the $\mathcal{G}_p(\im \omega_m)$ could cause large changes in the spectral function.

\subsection{Real-time Grassmann path integral}

The impurity partition function given by $\Zimp(t_f) = \Tr[\rhotot(t_f)]/\Tr[\rhobath]$, can be expressed as a real-time PI given by 
\begin{align}\label{eq:PI}
\Zimp(t_f) = \int \mathcal{D}[\boldabar,\bolda] \gK\left[\boldabar, \bolda \right]\prod_{p}\gI_{p}\left[\boldabar_{p}, \bolda_{p}\right],
\end{align}
where $t_f$ denotes the total evolution time, $\boldabar_p=\{\abar_p(t')\}$ and $\bolda_p=\{a_p(t')\}$ are Grassmann trajectories for flavor $p$ over the real-time interval $[0, t_f]$, $\boldabar = \{\boldabar_p, \boldabar_q, \cdots\}$ and $\bolda = \{\bolda_p, \bolda_q, \cdots\}$ are Grassmann trajectories for all flavors. 
The measure $\mea[\boldabar,\bolda]$ is given by  
\begin{align}
  \mea[\boldabar,\bolda]=\prod_{p, t'}\dd\abar_p(t')\dd a_p(t')e^{-\abar_p(t')a_p(t')}.
\end{align}
% $\gI[\boldabar_p,\bolda_p]$ denotes the IF for flavor $p$, which can be written as
The IF for flavor $p$, denoted by $\gI_p[\boldabar_p,\bolda_p]$, can be written as \begin{align}
  \label{eq:IF}
  \gI_p[\boldabar_p,\bolda_p]=e^{-\int_{\contour}\dd t'\int_{\contour}\dd t''\abar_p(t')\Delta(t',t'')a_p(t'')},
\end{align}
where $\contour$ denotes the Keldysh contour. The hybridization function $\Delta(t',t'')$ fully characterizes the effect of the bath and can be computed by  
\begin{align}\label{eq:hybrization-function}
\Delta(t',t'') = \im\mathcal{P}_{t', t''}\int \dd \omega J(\omega) D_{\omega}(t', t'').
\end{align}
Here $\mathcal{P}_{t', t''}=1$ if $t'$ and $t''$ are on the same Keldysh branch and $-1$ otherwise, $D_{\omega}(t', t'')$ is the free contour-ordered  Green’s function of the bath defined as
$\im D_{\omega}(t', t'') = \langle T_{\contour} \cop_{\omega}(t')\cdop_{\omega}(t'')\rangle_0$, and $J(\omega) = \sum_k V_k^2\delta(\omega - \varepsilon_k)$ is the bath spectrum density which ultimately determines the bath effects. Throughout this work, we consider the following bath spectrum density
\begin{align}
J(\omega) = \frac{\Gamma D}{2\pi}\sqrt{1 - \left(\frac{\omega}{D}\right)^2},
\end{align}
and set $\Gamma=0.1$, $D=2$ (these settings were also used in Refs.~\cite{BertrandWaintal2019,ThoennissAbanin2023b,ChenGuo2023} for studying non-equilibrium QIPs). $\Gamma$ will be used as the unit for the rest parameters.

For numerical evaluation of the PI in Eq.(\ref{eq:PI}), we discretize the time interval $[0, t_f]$ into $M$ discrete points with equal time step $\delta t = t_f/(M-1)$. Discretizing the Keldysh contour results in two branches: the forward branch ($+$) with Grassmann variables denoted as $a^+_{p, j}$ and $\abar^+_{p, j}$, and the backward branch ($-$) with GVs denoted as $a^-_{p, j}$ and $\abar^-_{p, j}$ ($j$ labels the discrete time step). We further denote $\bolda^{\pm}_p =  \{a^{\pm}_{p, M}, \cdots, a^{\pm}_{p, 1} \}$ and $\boldabar^{\pm}_p =  \{\abar^{\pm}_{p, M}, \cdots, \abar^{\pm}_{p, 1} \}$ as the discrete Grassmann trajectories, and denote $\boldabar^{\pm}_{, j}=\{\boldabar^{\pm}_{p, j}, \boldabar^{\pm}_{q, j}, \dots\}$ and $\bolda^{\pm}_{, j}=\{\bolda^{\pm}_{p, j}, \bolda^{\pm}_{q, j}, \dots\}$ as the set of GVs for all the flavors at discrete time step $j$. With these notations, the exponent in Eq.(\ref{eq:IF}) can be discretized via the QuaPI method~\cite{makarov1994-path,makri1995-numerical} as
\begin{align}\label{eq:quapi}
\int_{\contour}\dd t'\int_{\contour}\dd t''\abar_p(t')\Delta(t',t'')a_p(t'') \approx \sum_{\nu,\nu'}\sum_{j,k} \abar_{p, j}^{\nu} \Delta_{j,k}^{\nu,\nu'} a_{p,k}^{\nu'},
\end{align}
where $\nu,\nu'=\pm$ and
\begin{align}
\Delta_{j,k}^{\nu,\nu'} = \int_{j\delta t}^{(j+1)\delta t} dt' \int_{k\delta t}^{(k+1)\delta t} dt'' \Delta^{\nu,\nu'}(t',t'').
\end{align}
$\gK\left[\boldabar, \bolda \right]$ encodes the bare impurity dynamics which only depends on $\Himp$ and $\rhoimp$. In the discrete setting, it can be evaluated as
\begin{align}\label{eq:K}
  \gK[\boldabar,\bolda]=&\mel{-\bolda_{,M}}{\Uimp}{\bolda^+_{,M-1}}\cdots\mel{\bolda^+_{,2}}{\Uimp}{\bolda^+_{,1}} \nonumber \\
  &\times \mel{\bolda^+_{,1}}{\rhoimp}{\bolda^-_{,1}} \mel{\bolda^-_{,1}}{\Uimp^{\dagger}}{\bolda^-_{,2}} \times \cdots \nonumber \\
  &\times \mel{\bolda^-_{,M-1}}{\Uimp^{\dagger}}{\bolda_{,M}},
\end{align}
where we have used $\Uimp=e^{-\im\delta t \Himp}$. The term $\mel{\bolda^+_{,1}}{\rhoimp}{\bolda^-_{,1}}$ in Eq.(\ref{eq:K}) only depends on $\rhoimp$. The rest terms are propagators that are only determined by $\Himp$. 
% These propagators can either be calculated exactly for simple impurity models, or calculated in a numerically accurate \xs{exact?} way using the algorithm introduced in Ref.\cite{ChenGuo2024b}.
For the non-interacting case or the single-orbital Anderson impurity model, these propagators can be calculated analytically. For more complicated impurity models, they can be calculated in a numerically exact way using the algorithm introduced in Ref.~\cite{ChenGuo2024b}.

\subsection{Calculating non-equilibrium Green's functions}

Based on the discretized $\gI_{p}$ and $\gK$, one can compute the augmented density tensor (ADT) $\gA$ as (in practice $\gA$ is not built directly but only computed on the fly using a zipup algorithm~\cite{ChenGuo2023,ChenGuo2024b})
\begin{align}\label{eq:ADT}
\gA[\boldabar, \bolda] = \gK[\boldabar,\bolda]\prod_p \gI_p[\boldabar_p,\bolda_p].
\end{align}
Based on the ADT, one can straightforwardly calculate any Green's functions. For example, the discretized greater and lesser Green's functions in Eqs.(\ref{eq:neq_greater},\ref{eq:neq_lesser}) are found to be (assuming that $t = j\delta t$ and $t' = k\delta t$):
\begin{align}
\im G^{{\rm neq},>}_{p,q}(j,k) &= \Zimp^{-1}(t_f)\int \mea [\boldabar, \bolda] a_{p, j} \abar_{q, k} \gA[\boldabar, \bolda] ; \label{eq:neq_greater2} \\
\im G^{{\rm neq},<}_{p,q}(j, k) &= -\Zimp^{-1}(t_f)\int \mea [\boldabar, \bolda] \abar_{q, k} a_{p, j} \gA[\boldabar, \bolda]. \label{eq:neq_lesser2}
\end{align}
This means that the Green's functions are calculated by applying a product operator on the ADT and then integrating the resulting Grassmann MPS.

% In the following we will focus on $p=q=1$ and omit the flavor indices in the Green's functions.

\section{Impurity equilibration protocols}

In this section, we introduce the major techniques to calculate the equilibrium Green's functions from the real-time equilibration dynamics of the quantum impurity with a separable impurity-bath initial state. We also discuss the choices of impurity initial states. 

% which can be obtained by Fourier transformation of $G_{p,q}(t, t')$ into the frequency domain. 

\subsection{Calculating the equilibrium retarded Green's function}

\begin{figure}
  \includegraphics[width=\columnwidth]{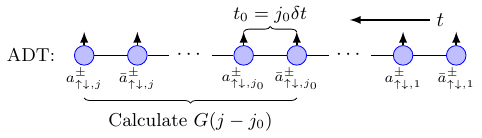} 
  \caption{Approximating the real-time equilibrium retarded Green's function $G(j-j_0)$ by the non-equilibrium retarded Green's function $G^{{\rm neq}}(j, j_0)$, where $t_0$ is the time taken for the impurity and bath to reach thermal equilibrium as a whole. The flavor indices are not shown. 
    }
    \label{fig:fig1}
\end{figure}

Our target is to calculate the equilibrium retarded Green's function $G_{p,q}(t,t')$, however only $G^{{\rm neq}}_{p,q}(j,k)$ can be directly calculated using the non-equilibrium GTEMPO method as in Eqs.(\ref{eq:neq_greater2}, \ref{eq:neq_lesser2}). Nevertheless, instead of obtaining the impurity-bath thermal state $\rhotot^{{\rm eq}}$ using imaginary-time evolution as in Eq.(\ref{eq:thermal_state}), one can also obtain it by evolving from the separable initial state $\rhotot^{{\rm neq}}$ for a long enough time, since the bath is infinite and the impurity will reach equilibrium with the bath eventually, independent of its initial state. In fact, this is a common practice in wave-function-based approaches~\cite{KohnSantoro2021,KohnSantoro2022,BauerTroyer2016}.
%\xs{We stress that the GTEMPO method is much more efficient in obtaining equilibration dynamics as compared to these wave-function-based approaches.}
% \textcolor{red}{The only possible issue here is that the bath degrees of freedom are integrated out in the GTEMPO method, which however, turns out not to prohibit us from utilizing the same idea in GTEMPO, as will be demonstrated in the following.}

We assume that after time $t_0$, the impurity equilibrates with the bath ($t_0$ will thus be referred to as the equilibration time), that is,
\begin{align}\label{eq:thermal_state2}
\rhotot^{{\rm eq}} \approx e^{-\im \Hop t_0} \rhotot^{{\rm neq}} e^{\im \Hop t_0}.
\end{align}
Substituting Eq.(\ref{eq:thermal_state2}) into Eq.(\ref{eq:greater}), we get 
\begin{align}\label{eq:greater2}
\im G^{>}_{p,q}(t, t') &\approx \Tr\left[\aop_p(t)\adop_q(t') e^{-\im \Hop t_0} \rhotot^{{\rm neq}} e^{\im \Hop t_0} \right]/\Tr[\rhotot^{\mathrm{neq}}(t_0)] \nonumber \\
&= \Tr\left[\aop_p(t+t_0)\adop_q(t' + t_0) \rhotot^{{\rm neq}} \right]/\Tr[\rhotot^{\mathrm{neq}}] \nonumber \\ 
&= G^{{\rm neq}, >}_{p,q}(t+t_0, t'+t_0),
\end{align}
where we have used the cyclic property of the trace. Similarly, we have for the lesser Green's function
\begin{align}\label{eq:lesser2}
G^{<}_{p,q}(t, t') \approx G^{{\rm neq}, <}_{p,q}(t+t_0, t'+t_0).
\end{align}
From Eqs.~(\ref{eq:greater2},\ref{eq:lesser2}), we can simply use Eqs.~(\ref{eq:neq_greater2}, \ref{eq:neq_lesser2}) to calculate the equilibrium retarded Green's functions. The only change is to compute those greater and lesser Green's functions between discrete times $ j + j_0$ and $k +j_0$ instead. This approach is illustrated in Fig.~\ref{fig:fig1}.
In practice, we find that a relatively small $t_0$ is sufficient for the impurity to reach equilibrium (See Sec.~\ref{sec:oneorbital}). In addition, to obtain $G_{p,q}(\omega)$ through the Fourier transform of $G_{p,q}(t-t')$, one needs to obtain $G_{p,q}(t, t')$ for infinitely large $t-t'$ in principle. A standard practice is to calculate $G_{p,q}(t, t')$ till a finite $t-t'$ and then extrapolate the results to infinity, for which we use the well-established linear prediction technique~\cite{WhiteAffleck2008,BarthelWhite2009}. In this work, we use the same hyperparameter settings for the linear prediction as in Ref.~\cite{WolfSchollwock2014b}.

\subsection{The impurity initial state}

\begin{figure}
  \includegraphics[width=0.8\columnwidth]{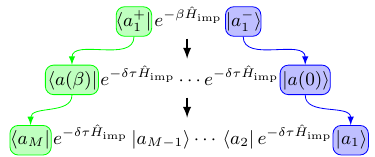} 
  \caption{Calculating the term $\mel{\bolda^+_{,1}}{\rhoimp}{\bolda^-_{,1}}$ in the expression of $\gK\left[\boldabar, \bolda \right]$, with $\rhoimp$ the impurity thermal state as in Eq.(\ref{eq:impurity_thermal_state}). $\delta\tau$ is a small imaginary-time step size such that each propagator $\langle a_{j+1} \vert e^{-\delta\tau \Himp}\vert a_j\rangle$ can be approximated by its first-order expression.
    }
    \label{fig:fig2}
\end{figure}

% One degree of freedom \xs{a bit strange} that can be explored in our method is the impurity initial state $\rhoimp$. 
Our real-time dynamics starts from a separable initial state of the impurity and bath.
In principle, for large enough $t_0$, the choice of the impurity initial state $\rhoimp$ does not matter. However, for numerical calculations, different choices of $\rhoimp$ may affect the quality of results due to finite equilibration time $t_0$. 
The effect of $\rhoimp$ is fully encoded in the term $\mel{\bolda^+_{,1}}{\rhoimp}{\bolda^-_{,1}}$ in the expression of $\gK$.
For simple states, this term can be calculated analytically. For example, for the vacuum state of a spinless fermion with $\rhoimp = \vert0\rangle\langle 0\vert$, we have
\begin{align}
\mel{a^+_{1}}{\rhoimp}{a^-_{1}} &= \mel{0}{e^{-\aop\abar_{1}^+}}{0}
       \mel{0}{e^{-a_{1}^-\adop}}{0} =1.
\end{align}

Other than the vacuum state,
another natural choice of $\rhoimp$ is the impurity thermal state, that is, 
\begin{align}\label{eq:impurity_thermal_state}
\rhoimp^{{\rm eq}} \propto e^{-\beta \Himp}. 
\end{align}
For general $\Himp$, $\mel{\bolda^+_{,1}}{\rhoimp^{{\rm eq}}}{\bolda^-_{,1}}$ can be calculated using the numerical algorithm shown in Fig.~\ref{fig:fig2}. The main idea is to split $\rhoimp^{{\rm eq}}$ as $ e^{-\beta \Himp} \approx e^{-\delta\tau \Himp} \cdots e^{-\delta\tau \Himp}$ with a small $\delta\tau$, which results in a GMPS after inserting GVs between the imaginary-time propagators (the same as the procedure used to build $\gK$ as a GMPS in the imaginary-time GTEMPO method~\cite{ChenGuo2024b}). Then one can easily integrate out all the intermediate GVs to obtain the final expression. This algorithm is very efficient since only the bare impurity dynamics is involved. Therefore one could set $\delta\tau$ to be arbitrarily small to obtain $\mel{\bolda^+_{,1}}{\rhoimp^{{\rm eq}}}{\bolda^-_{,1}}$ with high precision.
For simple impurities such as a single electron, $\mel{\bolda^+_{,1}}{\rhoimp^{{\rm eq}}}{\bolda^-_{,1}}$ may also be obtained analytically. Nevertheless, the above numeric algorithm is very convenient since it applies for general impurity problems and we found in practice that it is very numerically stable (Analytical expressions can easily run into numerical issues at low temperature if not taken with special care, since a very large $\beta$ appears in the matrix exponential).

\section{Results}
In this section, we demonstrate the accuracy and efficiency of our method with numerical examples. 
In our numerical calculations, we focus on the Green's function $G_{p,q}$ with $p=q$ and thus omit the flavor indices. We will use a maximum bond dimension $\chi$ for MPS bond truncation when building the MPS representation of the IF for each spin species, by keeping at most $\chi$ largest singular values at each bond (the same strategy as used in Ref.~\cite{ThoennissAbanin2023b}).

\subsection{The non-interacting case}
Generally, the dominant calculation in the GTEMPO method is the computation of Green's functions using the zipup algorithm (one can refer to Refs.~\cite{ChenGuo2023,ChenGuo2024b} for details of this algorithm). 
Concretely, if one represents the IF per flavor as a GMPS (referred to as the MPS-IF) with bond dimension $\chi$, then the computational cost for calculating one Green's function scales as $O(M\chi_{\gK}\chi^{n+1})$ for $n$ flavors~\cite{ChenGuo2024b} ($\chi_{\gK}$ is the bond dimension of the GMPS for $\gK$ which is usually a small constant). As a result, $\chi$ essentially determines the computational cost of the GTEMPO method. 

\begin{figure}
  \includegraphics[width=\columnwidth]{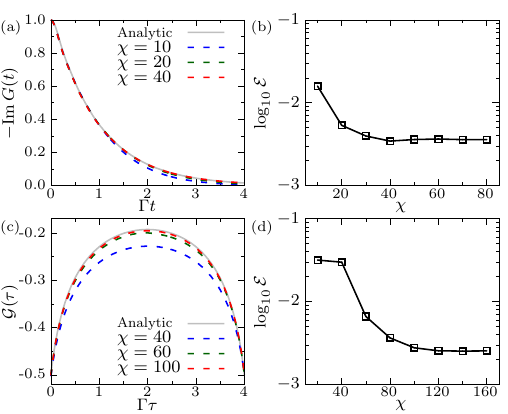} 
  \caption{
  (a) Imaginary part of the retarded Green's function for the Toulouse model as a function of time $t$ and (c) the Matsubara Green's function for the Toulouse model as a function of the imaginary time $\tau$, calculated at different values of $\chi$s. The gray solid lines in (a, c) are the analytical solutions. (b, d) The average error $\mse$ between the GTEMPO results and the analytical solutions as a function of $\chi$ calculated in (b) the real-time axis and (d) the imaginary-time axis. For this simulation we have used $\varepsilon_d=0$.
  For all the numerical simulations in this work, we have used $\delta t=0.05$ for the real-time GTEMPO method and $\delta\tau=0.1$ for the imaginary-time GTEMPO method. 
    }
    \label{fig:toulouse1}
\end{figure}

One major motivation for this work is that the bond dimension $\chi$ involved in the non-equilibrium GTEMPO method is usually much smaller than that in the imaginary-time GTEMPO method. We demonstrate such a difference in bond dimension growth in the non-interacting case with a single flavor. The corresponding non-interacting model is analytically solvable and referred to as the Toulouse model~\cite{leggett1987-dynamics} or the Fano-Anderson model~\cite{mahan2000-many}, with
\begin{align}
\Himp = \varepsilon_d \adop\aop .
\end{align}
We will use the square root of the mean square error to quantify the distance between two vectors $\vec{x}$ and $\vec{y}$, denoted as
\begin{align}
\mse(\vec{x}, \vec{y}) = \sqrt{||\vec{x} - \vec{y}||^2 / L},
\end{align}
where $L$ means the length of $\vec{x}$ and $|| \cdot ||$ means the Euclidean norm. We will refer to $\mse(\vec{x}, \vec{y})$ as the \textit{average error} afterwards and neglect the operands of $\mse$ when they are clear from the context. 
In Fig.~\ref{fig:toulouse1}(a,c), we show the imaginary part of the retarded Green's function calculated by the real-time GTEMPO method, and the Matsubara Green's function calculated by the imaginary-time GTEMPO method. The corresponding analytical solutions are shown in gray solid lines as comparisons. To clearly show the minimal bond dimensions required in these calculations, we plot the average errors of the real-time and imaginary-time GTEMPO results against the analytical solutions as functions of $\chi$ in Fig.~\ref{fig:toulouse1}(b,d). 
We can see that the average error saturates around $\chi=40$ in the real-time calculations, but only saturates around $\chi=100$ in the imaginary-time calculations, where the latter is $2.5$ times larger than the former. 
As a result, for one-orbital impurity models the computational cost of the latter would be $2.5^3/4 \approx 4$ times higher and for two-orbital impurity models $2.5^5/4 \approx 25$ times higher, where we have assumed that we evolve till $t=\beta$ in the real-time dynamics and we have already taken into account the fact that the number of GVs in the real-time GTEMPO method is two times larger than the imaginary-time GTEMPO method (generally speaking, the speedup is because that MPS -based algorithms are very sensitive to the bond dimension but only mildly depends on the system size~\cite{Schollwock2011}).

\begin{figure}
  \includegraphics[width=\columnwidth]{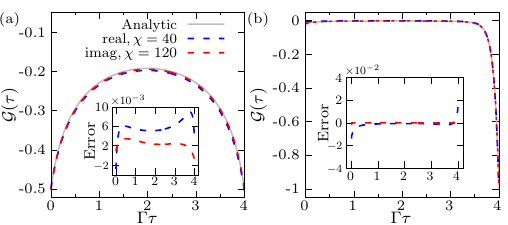} 
  \caption{The Matsubara Green's function of the Toulouse model for (a) $\varepsilon_d=0$ and (b) $\varepsilon_d=1$. In both panels, the gray solids lines are the analytical solutions, the red dashed lines are imaginary-time GTEMPO results calculated with $\chi=120$, the blue dashed lines are converted from the real-time GTEMPO results calculated with $\chi=40$. The insets in both panels show the difference between the GTEMPO results and the analytical solutions.
    }
    \label{fig:toulouse2}
\end{figure}

For the Toulouse model, it is easy to verify the GTEMPO results against the analytical solutions of the equilibrium Green's functions. For the single-orbital Anderson impurity model, the analytical solutions do not exist, therefore we verify the GTEMPO results against the state-of-the-art CTQMC results. Since the CTQMC impurity solver is only exact in the imaginary-time axis, we introduce an approach to benchmark the accuracy of our real-time calculations against imaginary-time calculations, that is, we use the spectral function obtained from $G(\omega)$ to calculate $\mathcal{G}(\tau)$, and then compare the obtained $\mathcal{G}(\tau)$ against those calculated using imaginary-time GTEMPO and CTQMC methods.
%obtain the spectral function, we perform the inverse of the analytical continuation to obtain $G(\tau)$ (which is numerically stable compared to the analytical continuation itself), which can then be directly compared to imaginary-time calculations.

%The detailed formalism for this conversion is shown in Appendix.~\ref{app:conversion}.

We present a proof of principle demonstration of the effectiveness of this comparison for the Toulouse model with $\varepsilon_d=0$ in Fig.~\ref{fig:toulouse2}(a) and $\varepsilon_d=1$ in Fig.~\ref{fig:toulouse2}(b). For the Toulouse model, the retarded Green's function is independent of the initial state \cite{mahan2000-many}, namely $G(t, t') = G^{{\rm neq}}(t, t')$. Therefore we simply choose $t_0=0$ in this case when calculating the equilibrium retarded Green's function. In both cases, we can see that the $\mathcal{G}(\tau)$ obtained by conversion from $G(t)$ matches very well with the analytical solution. 

\subsection{The single-orbital Anderson impurity model}\label{sec:oneorbital}

\begin{figure}
  \includegraphics[width=\columnwidth]{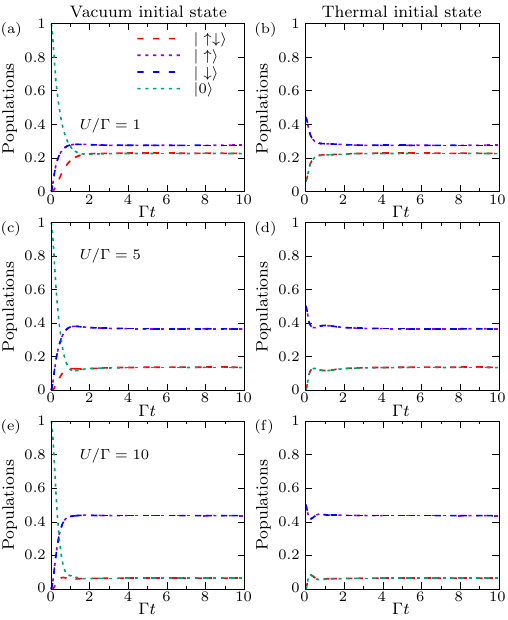} 
  \caption{Time evolution of the populations of the impurity for the single-orbital Anderson impurity model, where the impurity starts from the vacuum state for results in (a,c,e) and from the thermal state for results in (b,d,f). The three rows from top down are results for $U/\Gamma=1, 5, 10$ respectively. We have used $\chi=60$ for the MPS-IF per spin species in all the real-time GTEMPO calculations of the single-orbital AIM in this work. 
    }
    \label{fig:oneorbital1}
\end{figure}

Now we proceed to consider the single-orbital AIM with
\begin{align}
\Himp = \varepsilon_d \sum_{\sigma \in \{\uparrow, \downarrow\}} \adop_{\sigma}\aop_{\sigma} + U \adop_{\uparrow}\adop_{\downarrow}\aop_{\downarrow}\aop_{\uparrow},
\end{align}
where $\varepsilon_d$ is the on-site energy and $U$ is the interaction strength. 
We will focus on the half-filling case with $\varepsilon_d = -U/2$ which gives us an easy first verification of the results, we also set $\beta=40$ ($\Gamma\beta=4$) in all the following calculations.

We first access the quality of the approximate equilibrium state by studying the equilibration dynamics in Fig.~\ref{fig:oneorbital1}, where we plot the populations of the impurity in the four states: $\vert 0\rangle$ (no electron), $\vert \uparrow\rangle$ (spin up), $\vert\downarrow\rangle$ (spin down) and $\vert \uparrow\downarrow\rangle$ (double occupancy) as functions of time $t$. We show the results for impurity vacuum state in Fig.~\ref{fig:oneorbital1}(a,c,e) and results for impurity thermal state in Fig.~\ref{fig:oneorbital1}(b,d,f) respectively, for three different values of $U$ ($U/\Gamma = 1, 5, 10$). We can see that the populations in all these simulations have converged fairly well at $\Gamma t \approx 2$, and that the results for impurity thermal state (which approximately converge around $\Gamma t \approx 1$) clearly converge much faster than those for impurity vacuum state, which indicate that one could use a smaller equilibration time $t_0$ if starting from the impurity thermal state (which will significantly reduce the computational cost).

\begin{figure}
  \includegraphics[width=\columnwidth]{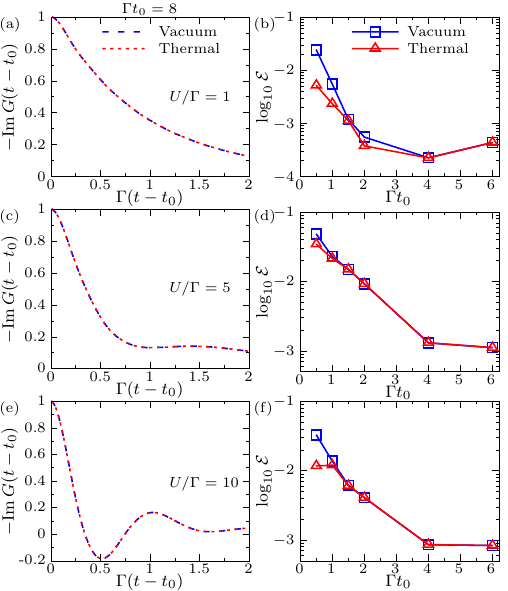} 
  \caption{(a,c,e) Imaginary part of the retarded Green's function as a function of $t-t_0$ with \gcc{$\Gamma t_0=8$} for (a) $U/\Gamma=1$, (c) $U/\Gamma=5$ and (e) $U/\Gamma=10$ respectively, where the blue and red dashed lines are results for impurity vacuum and thermal states. (b, d, f) The average error between the retarded Green's function $G^{{\rm neq}}(t, t_0)$ calculated with \gcc{$\Gamma t_0=8$} and the $G^{{\rm neq}}(t, t_0)$ calculated with smaller $t_0$ for (b) $U/\Gamma=1$, (d) $U/\Gamma=5$ and (f) $U/\Gamma=10$ respectively.
    }
    \label{fig:oneorbital2}
\end{figure}

In Fig.~\ref{fig:oneorbital2}, we further quantify the error occurred in the retarded Green's function by using different values of $t_0$. Concretely, in Fig.~\ref{fig:oneorbital2}(a,c,e), we show $G^{{\rm neq}}(t, t_0)$ calculated with \gcc{$\Gamma t_0 = 8$} for $U/\Gamma = 1, 5, 10$ respectively, for both the impurity vacuum state (blue dashed lines) and the impurity thermal state (red dashed lines). We can see that these two lines are completely on top of each other (the average errors are less than \gcc{$10^{-5}$} for all $U$s). Then in Fig.~\ref{fig:oneorbital2}(b,d,f), we take $G^{{\rm neq}}(t, t_0)$ calculated with \gcc{$\Gamma t_0 = 8$} as the baseline (which has well reached equilibrium for all the cases we have considered from Fig.~\ref{fig:oneorbital1}), and then compute the average error between it and the $G^{{\rm neq}}(t, t_0)$ calculated with smaller $t_0$. We can see that the average errors quickly decrease to less than $10^{-2}$ when $t_0$ reaches $\Gamma t_0=2$ \gcc{and saturates at around $\Gamma t_0=6$ for both initial states (which is likely due to the first-order time discretization error in Eq.(\ref{eq:quapi}))}. We can also clearly see that the results for impurity thermal state converge much faster than those for impurity vacuum state.

% In Fig.~\ref{fig:oneorbital1}, we study the effect of the equilibrium time $t_0$ and the choice of impurity initial state on the quality of the equilibrium retarded Green's functions for $U/\Gamma=5$. The effects of $t_0$ and the impurity initial state are shown from $4$ dimensions: 1) In Fig.~\ref{fig:oneorbital1}(a), we show the average fermionic occupation
% \begin{align}
% n(t) = \Tr[\adop_{\uparrow}(t)\aop_{\uparrow}(t)\rhotot^{{\rm neq}}] = \Tr[\adop_{\downarrow}(t)\aop_{\downarrow}(t)\rhotot^{{\rm neq}}]
% \end{align}
% as a function of $t$, which should approach $n(\infty)=0.5$ asymptotically. We can see that this holds for impurity vacuum state after $\Gamma t\approx 1$, while holds for any $t$ for impurity thermal state; 2) In Fig.~\ref{fig:oneorbital1}(b), we show the imaginary part of the retarded Green's $G^{{\rm neq}}(t, t_0)$ as a function of $t-t_0$ for different values of $t_0$ (for the symmetric spectral function we have ${\rm Re} \left[G(t)\right]=0$), which should be independent of $t_0$ if the impurity and bath have well-reached equilibrium. We can see that the two lines for $\Gamma t_0=1, 1.5$ are already nicely on top of each other; 3) 

\begin{figure}
  \includegraphics[width=\columnwidth]{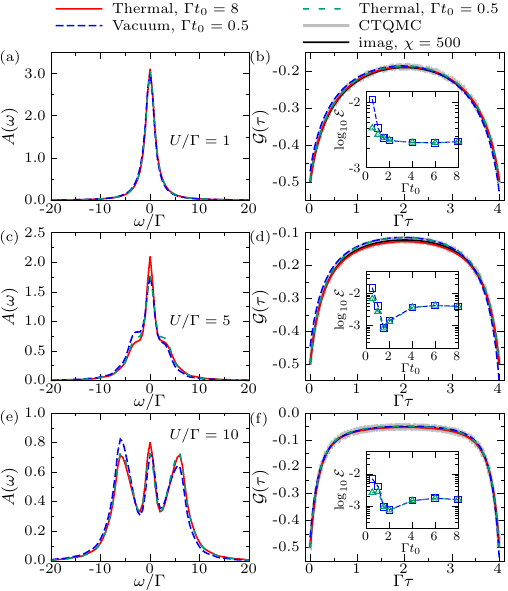} 
  \caption{(a,c,e) The spectral function $A(\omega)$ as a function of $\omega$ for (a) $U/\Gamma = 1$, (c) $U/\Gamma = 5$ and (e) $U/\Gamma = 10$ respectively, where the red solid lines are results calculated with \gcc{$\Gamma t_0=8$} for the impurity thermal state, the blue and green dashed lines are results calculated with $\Gamma t_0=0.5$ for the impurity vacuum and thermal states respectively. (b,d,f) The Matsubara Green's function $\mathcal{G}(\tau)$ converted from the $A(\omega)$ calculated in (a,c,e) respectively. \gcc{The black and gray solid lines in (b,d,f) are the imaginary-time GTEMPO results and CTQMC results. The insets in (b,d,f) show the average errors between $\mathcal{G}(\tau)$ converted from the $A(\omega)$ calculated with different $t_0$ and the imaginary-time GTEMPO results, where the blue dashed line with square and green dashed line with triangle are for the impurity vacuum and thermal states respectively.} }
    \label{fig:oneorbital3}
\end{figure}

As application, we plot the spectral function $A(\omega) $ as a function of $\omega$, obtained using different values of $t_0$, in Fig.~\ref{fig:oneorbital3}(a,c,e) for $U/\Gamma = 1,5,10$ respectively, where the red solid lines are results calculated with \gcc{$\Gamma t_0=8$} for impurity thermal state, while the blue and green dashed lines are results calculated with $\Gamma t_0=0.5$ for impurity vacuum and thermal states respectively. In Fig.~\ref{fig:oneorbital3}(b,d,f), we plot $\mathcal{G}(\tau)$ converted from the $A(\omega)$ calculated in Fig.~\ref{fig:oneorbital3}(a,c,e) respectively, where we have also shown the CTQMC results and the imaginary-time GTEMPO results calculated with $\chi=500$ as comparisons.
\gcc{To reduce the finite-time discretization error in this conversion (since we need to perform the Fourier transformation of $G(t)$ in Eq.(\ref{eq:fourier})) we have used a simple linear interpolation scheme to obtain refined real-time data with a smaller time step size $10^{-4}/\Gamma$.}
We can see that the $\mathcal{G}(\tau)$ converted from $A(\omega)$ calculated with \gcc{$\Gamma t_0=8$} well agrees with the CTQMC and imaginary-time GTEMPO results. Interestingly, the $A(\omega)$ calculated with $\Gamma t_0 = 0.5$ for impurity thermal state agrees fairly well with the $A(\omega)$ calculated with \gcc{$\Gamma t_0 = 8$} (similar for the corresponding $\mathcal{G}(\tau)$), while in comparison the $A(\omega)$ calculated with $\Gamma t_0 = 0.5$ for impurity vacuum state is very different from that calculated with \gcc{$\Gamma t_0=8$}. \gcc{The insets in Fig.~\ref{fig:oneorbital3}(b,d,f) show the average errors between $\mathcal{G}(\tau)$ converted from the $A(\omega)$ calculated with different $t_0$ and the imaginary-time GTEMPO results, which decrease at the beginning and saturates at $\Gamma t_0\geq 4$ (there is a slight increase of average error for $U/\Gamma=5,10$ when $\Gamma t_0$ increases from $1.5$ to $4$, which may be a coincidence due to the errors occurred during the conversion from $A(\omega)$ to $\mathcal{G}(\tau)$).}
With these results, we can see that the equilibrium retarded Green's function can indeed be accurately calculated using our method. Moreover, by preparing the initial state of the impurity in a local thermal state, the equilibration time $t_0$ can be significantly shortened.

% In Fig.~\ref{fig:oneorbital1}(c, e), we show the spectral density $A(\omega) $ as a function of $\omega$, obtained using different values of $t_0$. We can see that for $\Gamma t_0=0.5$ with impurity vacuum state, the spectral density is not symmetry around $\omega=0$ which is qualitatively wrong; 4) 

% In Fig.~\ref{fig:oneorbital1}(d, f), we show $G(\tau)$ converted from $A(\omega)$ obtained with different values of $t_0$, where the CTQMC results and the imaginary-time GTEMPO results calculated with $\varsigma=10^{-10}$ are also shown as comparison. We can see that in the middle regime, the real-time results obtained with different $t_0$ are all fairly accurate. While at the boundaries, it is clear the real-time results with $\Gamma t_0=0.5$ with impurity vacuum state are qualitatively wrong.
% Therefore, we can see that with a large enough $\Gamma t_0 \geq 1$, we can indeed obtain accurate results with different impurity initial states, while with the impurity thermal state, we can already obtain very accurate results with $\Gamma t_0=0.5$.

% To demonstrate the general effectiveness of our method, we also show the results for interaction strengths $U/\Gamma=1,10$ in Fig.~\ref{fig:oneorbital2}, with $\rhoimp=\rhoimp^{{\rm eq}}$. We can see that with the impurity thermal state, our method already gives very accurate results at $\Gamma t_0=0.5$ for the wide range of interaction strengths we have considered.

\section{Conclusion}
In summary, we have proposed a real-time impurity solver based on the non-equilibrium Grassmann time-evolving matrix product operators method. We evolve the impurity model in real time from a separable initial state, and after a long enough equilibration time the impurity model would reach the equilibrium. Then the equilibrium retarded Green's function together with the spectral function can be calculated. Our approach only contains three hyperparameters: the time discretization, the maximum MPS bond dimension and the equilibration time. We demonstrate the performance advantage of this method against the imaginary-time GTEMPO method by showing that the Grassmann MPS generated for the influence functional in the real-time calculations has a much smaller bond dimension compared to that in imaginary-time calculations. We also demonstrate the effectiveness of this method for the single-orbital Anderson impurity model for a wide range of interacting strengths and show that by starting from a thermal initial state of the impurity, one can obtain an accurate spectral function even with a relatively small equilibrium time. 
Our method thus opens the door to using the GTEMPO method as a real-time impurity solver for DMFT. 

% We aim to address two problems in our future works. First, we will focus on improving the efficiency further and apply our method to multi-orbital models. Second, we aim to integrate our GTEMPO-based real-time impurity solver into a complete DMFT loop to study to realistic strongly correlated materials.

\acknowledgements 
The CTQMC calculations in this work are done using the TRIQS package~\cite{ParcolletSeth2015,SethParcollet2016}.
This work is supported by the National Natural Science Foundation of China under Grant No. 12104328 and 12305049. C. G. is supported by the Open Research Fund from the State Key Laboratory of High Performance Computing of China (Grant No. 202201-00).

%\bibliographystyle{apsrev4-2}
% \bibliography{refs}
%apsrev4-2.bst 2019-01-14 (MD) hand-edited version of apsrev4-1.bst
%Control: key (0)
%Control: author (8) initials jnrlst
%Control: editor formatted (1) identically to author
%Control: production of article title (0) allowed
%Control: page (0) single
%Control: year (1) truncated
%Control: production of eprint (0) enabled
%

\end{document}